\documentclass[11pt]{article}
\begin{document}

\title{Null particle solutions in three-dimensional (anti-) de~Sitter spaces}

\author{Rong-Gen Cai\thanks{E--mail: {\tt cai@wormhole.snu.ac.kr}}
\\
\\ Center for Theoretical Physics, \\
Seoul National University,\\
Seoul 151-742, Korea\\
\\
and J. B. Griffiths\thanks{E--mail: {\tt J.B.Griffiths@Lboro.ac.uk}} \\ \\
Department of Mathematical Sciences \\
Loughborough University \\
Loughborough, Leics. LE11 3TU, U.K. \\ }

\maketitle

\begin{abstract}
We obtain a class of exact solutions representing null particles moving in
three-dimensional (anti-) de~Sitter spaces by boosting the corresponding
static point source solutions given by Deser and Jackiw. In de~Sitter space
the resulting solution describes two null particles moving on the (circular)
cosmological horizon, while in anti-de~Sitter space it describes a single null
particle propagating from one side of the universe to the other. We also boost
the BTZ black hole solution to the ultrarelativistic limit and obtain the
solution for a spinning null particle moving in anti-de~Sitter space. We find
that the ultrarelativistic geometry of the black hole is exactly the same as
that resulting from boosting the Deser-Jackiw solution when the angular
momentum of the hole vanishes. A general class of solutions is also obtained
which represents several null particles propagating in the Deser-Jackiw
background. The differences between the three-dimensional and four-dimensional
cases are also discussed. 
\end{abstract}

\newpage

\section{Introduction}

Although the Einstein equations still hold in three-dimensional spacetime,
the nature of gravity is quite different from that in four-dimensional
spacetime. Because the Einstein and Riemann tensors are equivalent in
three-dimensional spacetime, general relativity is dynamically trivial there.
That is, the vacuum spacetime is flat. The localised sources have effects only
on the  global geometry. In 1984, Deser, Jackiw, and 't Hooft \cite{DJH1}
investigated in detail the Einstein gravity with static point sources in
three-dimensional spacetime. For a single static particle, the geometry is
given by cutting a sector out of the Euclidean two-plane along two straight
lines, and identifying the edges to  form a cone. Gravity theories with
lightlike sources and spacelike source in three-dimensional flat spacetime
have also been analysed in \cite{DS} and \cite{DJH2}, respectively.

When a nonvanishing cosmological constant is introduced to the
three-dimensional Einstein gravity, some significant changes occur. In this
case, the spacetime has constant curvature and corresponds either to de~Sitter
or to anti-de~Sitter space. In the de~Sitter space, the static two-particle
solution is a sphere minus a wedge with the edges identified.  This is because
the two-space is a sphere in a three-dimensional covering space. To obtain the
effect of a point particle one can cut the sphere from the location of the
source along two great circles. On a sphere, these cuts meet again at the
antipodal point. By identifying along the cuts, this procedure automatically
creates a ``mirror'' source. There is no pure one-particle solution globally.
For the anti-de~Sitter case, the two-space is a hyperboloid. This can be cut
along two lines and the cuts identified to produce single particle solutions.
Deser and Jackiw \cite{DJ} have obtained a metric  (hereafter denoted the DJ
solution) and confirmed the above geometrical picture for static point sources
by directly solving the Einstein equations with a cosmological constant.

In this paper we investigate the Einstein gravity with null particle sources
in the three-dimensional de~Sitter and anti-de~Sitter spaces. Initially, we
employ the boost method that was first used by Aichelburg and Sexl \cite{AS}
to derive the gravitational field generated by a photon. By boosting the
Schwarzschild solution to the ultrarelativistic limit in which the velocity of
the source approaches the speed of light and the mass is scaled to zero in an
appropriate manner, Aichelburg and Sexl derived a solution describing an
impulsive gravitational wave propagating in a flat spacetime. This method has
subsequently been widely used to investigate the gravitational fields
generated by various null sources moving in flat spacetimes (for a brief
review see \cite{CJS} and references cited therein). Due to the fact that the
four-dimensional (anti-) de~Sitter space can be represented as a
four-dimensional hyperboloid embedded in a five-dimensional flat spacetime,
Hotta and Tanaka \cite{HT} succeeded in obtaining exact solutions for null
particles moving in (anti-) de~Sitter spacetimes by boosting the
Schwarzschild--(anti-)de~Sitter solutions. The impulsive wave surfaces
generated have been discussed in detail by Podolsk\'y and Griffiths \cite{PG1}.
Further they considered more general gravitational wave solutions in
(anti-) de~Sitter spaces \cite{PG2}. These can be interpreted as impulsive
gravitational waves generated by an arbitrary distribution of null particles
each with arbitrary multipole structure. 

The plan of this paper is as follows. In the next section, we will introduce 
the DJ solution and  boost the spacetime to the ultrarelativistic limit in the
three-dimensional de~Sitter and anti-de~Sitter spaces and then analyse the
resulting geometries. We will also boost the Ba\~nados--Teitelboim--Zanelli
(BTZ) black hole \cite{BTZ} in the anti-de~Sitter space in section~3. Although
the BTZ black hole solution is quite different from the DJ solution globally,
we find that, when the angular momentum of the BTZ black hole vanishes, the
resulting geometries are equivalent to each other. In section~4 we will
consider the null-particle solution in the DJ background, and further confirm
the result derived using the boost method. A brief discussion of the main
results is included in section~5.

\section{Boosting the DJ solutions in the (anti-) de~Sitter spaces}

The three-dimensional Einstein equations with a nonvanishing cosmological
constant can be written as
\begin{equation}
\label{eq}
R_{\mu\nu} -{\textstyle{1\over2}}R\,g_{\mu\nu} +\Lambda\,g_{\mu\nu}
=8\pi\,T_{\mu\nu},
\end{equation}
 where $\Lambda$ denotes the cosmological constant and $T_{\mu\nu}$ the
energy-momentum tensor of the sources. Here the gravitational constant
$G$ has been set to one.

The solutions which describe a static point particle at the origin 
in the (anti-) de~Sitter spaces were found to be \cite{DJ}
\begin{equation}
\label{DJ}
ds^2=-N^2(R)dt^2 +\Phi(R)(dR^2+R^2d\phi^2),
\end{equation}
where
\begin{eqnarray}
&& \Phi(R)=\frac{4\alpha^2}{\Lambda R^2[(R/R_0)^{\alpha}+
    (R/R_0)^{-\alpha}]^2}, \nonumber\\
&& N(R)=\frac{(R/R_0)^{\alpha}-(R/R_0)^{-\alpha}}
        {(R/R_0)^{\alpha} +(R/R_0)^{-\alpha}},
\end{eqnarray}
$R_0$ is an integration constant and $\alpha =1-4M$. The constant $M$ is the 
mass of the point particle. Performing a simple coordinate transformation in
 (\ref{DJ}) gives 
\begin{equation}
\label{dj}
ds^2=-(1-\Lambda r^2/\alpha^2) dt^2 +
     \alpha^{-2}(1-\Lambda r^2/\alpha^2)^{-1}dr^2 +r^2d\phi^2.
\end{equation}
When $\Lambda >0$, the solution (\ref{dj}) has a cosmological event horizon at
$r_c=\alpha/\sqrt{\Lambda}$ with surface gravity $\kappa=\sqrt{\Lambda}$. When
$\alpha =1$, that is for the vacuum case $M=0$, the DJ solution (\ref{dj})
reduces to the familiar form
\begin{equation}
\label{vacuum}
ds^2=-(1-\Lambda r^2)dt^2 +(1-\Lambda r^2)^{-1}dr^2 +r^2d\phi^2,
\end{equation}
which is just the three-dimensional de~Sitter ($\Lambda >0$) or anti-de~Sitter
($\Lambda <0$) space in static coordinates.
          
Similar to the case in four dimensions, the three-dimensional (anti-) 
de~Sitter space can also be represented as a hyperboloid embedded in a
four-dimensional flat spacetime. Let us first consider the case of the
de~Sitter space. 

\bigskip
(i). {\it In de~Sitter space.} In this case, the hyperboloid
satisfies
\begin{equation}
\label{decon}
-Z_0^2 +Z_1^2+Z_2^2+Z_3^2=a^2,
\end{equation}
where $a^2=1/\Lambda$. The de~Sitter space  can be expressed as the following
$SO(1,3)$ invariant line element satisfying the constraint (\ref{decon}) 
\begin{equation}
\label{desitter}
ds^2_{\rm ds}=-dZ_0^2 +dZ^2_1+dZ_2^2+dZ_3^2.
\end{equation}
Obviously, when we parametrize the hypersurface (\ref{decon}) with the
following coordinates
\begin{eqnarray}
\label{depara}
&& Z_0=\sqrt{a^2-r^2}\sinh(t/a), \qquad Z_1=r\cos\phi,
 \nonumber \\
&& Z_3=\pm \sqrt{a^2-r^2}\cosh(t/a), \qquad Z_2=r\sin\phi,
\end{eqnarray}
the metric (\ref{vacuum}) can be deduced from (\ref{desitter}). When boosting
the DJ solution in the de~Sitter space (\ref{desitter}), it is appropriate
first to expand the solution (\ref{dj}) up to the first order of the mass
$M$ (higher order contributions will vanish due to the boost). This yields
\begin{equation}
\label{defirst1}
ds^2 \approx ds^2_{\rm ds} +8M\Lambda r^2 dt^2 
+\frac{8M}{(1-\Lambda r^2)^2}dr^2,
\end{equation}
where $ds^2_{\rm ds}$ denotes the de~Sitter space (\ref{vacuum}). Using
the coordinates (\ref{depara}), we can rewrite (\ref{defirst1}) as
\begin{eqnarray}
\label{defirst2}
ds^2 &=& ds^2_{\rm ds}+\frac{8M}{(Z_3^2-Z_0^2)^2}\left [(a^2+Z_0^2-Z_3^2)
   (Z_3dZ_0-Z_0dZ_3)^2 \right. \nonumber \\
   & & +\left. \frac{a^4}{(a^2+Z_0^2+Z_3^2)}(Z_3dZ_3-Z_0dZ_0)^2 \right].
\end{eqnarray}
We now make a Lorentz boost in the $Z_1$-direction, that is, a Lorentz
transformation
\begin{equation}
\label{lorentz}
Z_0 \rightarrow \frac{Z_0+vZ_1}{\sqrt{1-v^2}},\ \ Z_1\rightarrow 
\frac{vZ_0+Z_1}{\sqrt{1-v^2}},\ \ Z_2\rightarrow Z_2, \ \ Z_3\rightarrow Z_3,
\end{equation}
where $v$ is the boost velocity. To obtain a result of physical interest,
the mass must be reduced to zero in an appropriate way. Following \cite{AS},
we scale mass as
\begin{equation}
\label{scale}
M=p\sqrt{1-v^2},
\end{equation}
where $p$ is a constant which can be interpreted as the energy of the null
particle. Substituting (\ref{lorentz}) and (\ref{scale}) into
(\ref{defirst2}), we obtain
\begin{eqnarray}
\label{defirst3}
ds^2&=& ds^2_{\rm ds} +\frac{8p\sqrt{1-v^2}}{(Z_3^2-z^2)^2}\left [
       (a^2+z^2-Z_3^2)(Z_3dz-zdZ_3)^2 \right. \nonumber\\
     & & \left. +\frac{a^4}{(a^2+z^2-Z_3)}(Z_3dZ_3-zdz)^2\right],
\end{eqnarray}
where $z^2=(Z_0+vZ_1)^2/(1-v^2)$. Using the identity
\begin{equation}
\lim_{v\rightarrow 1}\frac{1}{\sqrt{1-v^2}}f(z^2) =\delta(Z_0+Z_1)
  \int^{\infty}_{-\infty}f(z^2)dz,
\end{equation}
and taking the limit $v\rightarrow 1$ in (\ref{defirst3}), we obtain
\begin{equation}
\label{deshock}
ds^2=ds^2_{\rm ds}-8\pi p|Z_2|\delta(Z_0+Z_1) (dZ_0+dZ_1)^2.
\end{equation}
 This looks like an impulsive wave solution in the de~Sitter space, located on
the surface $Z_0+Z_1=0$, $Z_2^2+Z_3^2=a^2$ which at any time is a circle of
constant radius.

In order to further analyse this solution, it proves convenient to use the
following coordinates~\cite{PG1}
\begin{eqnarray}
\label{decon2}
&& Z_0=\frac{1}{2\eta}[a^2-\eta^2 +(x-a)^2 +y^2], \qquad
Z_1=\frac{a}{\eta}(x-a),
   \nonumber\\
&& Z_3=\frac{1}{2\eta}[a^2+\eta^2-(x-a)^2-y^2], \qquad Z_2=\frac{a}{\eta}y.
\end{eqnarray}
 Further we can put
 \begin{equation}
 x=\rho\cos\phi, \qquad y=\rho\sin\phi,
 \end{equation}
 with $\rho \in [0,\infty)$, $\phi \in [0,2\pi)$. The de~Sitter space can then
be described as
 \begin{equation}
 ds^2_{\rm ds}=\frac{a^2}{\eta^2} (-d\eta^2 +d\rho^2 +\rho^2d\phi^2),
 \end{equation}
 which is in conformally flat form. The solution (\ref{deshock}) can then be
rewritten as 
 \begin{equation}
 \label{decircle}
 ds^2=ds^2_{\rm ds}-8\pi p\,a |\sin\phi| \big[\delta(\eta-\rho)(d\eta-d\rho)^2
   +\delta (\eta +\rho)(d\eta+d\rho)^2\big].
 \end{equation}
 This looks like two impulsive wavefronts. However, as pointed out in
\cite{PG1} in the four-dimensional case, both components are required for the
conformal picture to be geodesically complete. Because $\rho\ge0$, the term
$\delta(\eta-\rho)$ does not vanish for $\eta\ge0$ only, while
$\delta(\eta+\rho)$ is required for $\eta\le0$. From (\ref{decircle}), it is
clear that the particles are located on the  circle $\rho=|\eta|$ which is the
cosmological horizon of the de~Sitter space.

It can then be shown that the energy-momentum tensor is only non-zero at the
two points $Z_0+Z_1=0$, $Z_2=0$, $Z_3=\pm a$ which thus represent two null
particles. At all other points on this null surface, the impulsive component
can in fact be removed by a discontinuous coordinate transformation. The
solution can thus be represented as a three-dimensional de~Sitter space cut
along the cosmological horizon $Z_0+Z_1=0$, with the two halves reattached in
such a way as to create two null particles at the points $Z_2=0$, or $y=0$, or
$\phi=0,\pi$, which are at opposite points on the horizon. This situation is
very like that in the four-dimensional case, in which instead of the circle
the wave surface is spherical and the particles are located at opposite poles.
The significant difference, however, is that in the four-dimensional case the
Weyl tensor has some non-zero components on the spherical surface and these 
can be interpreted as describing gravitational wave components generated by
the null particles. In the three-dimensional theory such free gravitational
waves cannot occur.

\bigskip
 (ii). {\it In anti-de~Sitter space}. We now turn to the case of the
three-dimensional anti-de~Sitter space. This can be regarded as a hyperboloid
 \begin{equation}
 \label{andcon}
 -Z_0^2+Z_1^2+Z_2^2-Z_3^2=-a^2,
 \end{equation}
 embedded in an $SO(2,2)$ invariant four-dimensional flat spacetime
 \begin{equation}
 \label{ads}
 ds^2_{\rm ads}=-dZ_0^2 +dZ_1^2 +dZ_2^2 -dZ_3^2,
 \end{equation}
 where $a^2=-1/\Lambda >0$. Obviously, the anti-de~Sitter space (\ref{vacuum})
 can be parametrized by the following coordinates
 \begin{eqnarray}
 \label{adsco}
 && Z_0=\sqrt{a^2+r^2}\sin(t/a), \qquad  Z_1= r\cos\phi, \nonumber \\
 && Z_3=\sqrt{a^2+r^2}\cos(t/a), \qquad Z_2=r\sin\phi.
 \end{eqnarray}
We now boost the DJ solution in the anti-de~Sitter space. Again expanding the
solution up to the first order in the mass $M$ and using the coordinates
(\ref{adsco}), we arrive at
 \begin{eqnarray}
 \label{adsfirst1}
 ds^2 &=& ds^2_{\rm ads} +\frac{8M}{(Z_0^2+Z_3^2)^2} \left[ (Z_0^2 +Z_3^2 -a^2)
          (Z_3dZ_0-Z_0dZ_3)^2 \right. \nonumber \\
     & & \left. +\frac{a^4}{(Z_0^2+Z_3^2-a^2)}(Z_3dZ_3-Z_0dZ_0)^2\right].
\end{eqnarray} 
Repeating the same steps as in the case of the de~Sitter space, that is,
using the Lorentz transformation (\ref{lorentz}), rescaling the mass as
(\ref{scale}), and taking  the limit $v\rightarrow 1$, finally we can obtain
\begin{equation}
\label{andshock}
ds^2=ds^2_{\rm ads}-8\pi p|Z_2|\delta(Z_0+Z_1)(dZ_0+dZ_1)^2.
\end{equation}
Comparing with (\ref{deshock}), it is easy to see that the expression for the
apparent impulsive part is the same as that in the de~Sitter space. However,
the interpretation is quite different. In this case the impulsive component is
given by the surface $Z_0+Z_1=0$, $Z_2^2-Z_3^2=-a^2$ which at any time is a
hyperbola. Let us now analyse this solution.
 
 In the anti-de~Sitter space, introduce first the following coordinates
 \begin{eqnarray}
 \label{anticon2}
 && Z_0=\frac{1}{2x}[a^2-\eta^2 +x^2 +(y-a)^2], \qquad Z_1=\frac{a}{x}(y-a),
   \nonumber \\
 && Z_2= \frac{1}{2x}[a^2+\eta^2 -x^2-(y-a)^2], \qquad Z_3=\frac{a}{x}\eta,
 \end{eqnarray}
 and then use
 \begin{equation}
 \label{con}
 x=\rho \cos\phi, \ \ y=\rho\sin\phi.
 \end{equation}
This produces the anti-de~Sitter space written in the conformally flat form 
 \begin{equation}
 \label{adscf}
 ds^2_{\rm ads}=\frac{a^2}{\rho^2\cos^2\phi}[-d\eta^2+d\rho^2 +\rho^2
    d\phi^2].
 \end{equation}
 The solution (\ref{andshock}) can be rewritten in these coordinates as
 \begin{equation}
 \label{boloid}
 ds^2=ds^2_{\rm ads}-\frac{8\pi p\ a |\sin\phi|}{\cos^2\phi}
     [\delta(\eta-\rho)(d\eta-d\rho)^2 +\delta(\eta+\rho)(d\eta+d\rho)^2].
 \end{equation}
 Here it should be stressed that the solution (\ref{boloid}) does not mean two
impulses again. As in the de~Sitter case, $\delta (\eta -\rho)$ works only for
$\eta >0$ while $\delta(\eta+\rho)$ for $\eta <0$. The two components are
required for globally geodesic completeness. From (\ref{boloid}), it is clear
that the impulsive component is located on the line $\rho=|\eta|$, or
$x^2+y^2=\eta^2$. However, this is not a circle --- according to (\ref{adscf})
it is conformal to a circle, and the coordinate $\phi$ is restricted to
$-\pi/2<\phi<\pi/2$. In fact it is a hyperbola. It can then be shown that this
solution represents a single null particle located at $Z_2=0$ on the null
surface $Z_0+Z_1=0$, i.e. at $y=0$, $x=\eta$ (or $x=-\eta$). The particle thus
clearly propagates from one side of the universe to the other and (since this
spacetime contains closed timelike lines) may then be considered to propagate
back in the opposite direction. 

\bigskip
Thus, by boosting the DJ solutions, we have obtained two kinds of exact
solutions describing null particles moving in the three-dimensional de~Sitter
and anti-de~Sitter spaces. (In section 4 we will further confirm these results
by directly solving the Einstein equations.) Although static particles only
have effect on the global geometry, which is quite different from the
situation in the four-dimensional case, we still find that the boost method is
sufficiently powerful to derive null particle solutions from their
corresponding static particle solutions. In the next section we will boost the
BTZ black hole solution in the anti-de~Sitter space. In the static situation,
this is quite different from the DJ solution from the aspect of global
properties. However, the resulting ultrarelativistic geometry is found to be
identical, at least in the non-rotating case.

\section{Boosting the BTZ black hole solution in the anti-de~Sitter
space}         
 
 Due to the special properties of three-dimensional gravity,
 it was a surprising discovery when Ba\~nados, Teitelboim,
 and Zanelli \cite{BTZ} claimed that they found a black hole solution in the
 Einstein gravity with a negative cosmological constant. The solution they
 found is
 \begin{equation}
 \label{btz}
 ds^2=-N^2(r)dt^2+N^{-2}dr^2+r^2(N^{\phi}(r)dt +d\phi)^2,
 \end{equation}
 where
 \begin{equation}
  N^2=-8M +\frac{r^2}{a^2} +\frac{16J^2}{r^2}, \ \
  N^{\phi}=-\frac{4J}{r^2}.
 \end{equation}
 Here $-1/a^2 $ denotes the negative cosmological constant. The integration
constants $M$ and $J$ can be interpreted as the mass and angular momentum of
the black hole. This black hole has two horizons at
 \begin{equation}
 r_{\pm}^2=4Ma^2\left[ 1\pm \sqrt{1-\left(\frac{J}{Ma}\right)^2}\right],
 \end{equation}
 provided $M>0$ and $J<Ma$. This solution is asymptotically an anti-de~Sitter
spacetime and can be constructed by identifying some discrete points in the
three-dimensional anti-de~Sitter space. It is of interest to note, however,
that when the mass and angular momentum of the hole vanish, the solution does
not reduce to the anti-de~Sitter space. Rather, the anti-de~Sitter spacetime
(\ref{vacuum}) can only be obtained from (\ref{btz}) in the limit as $8M\to-1$
and $J\to0$. 
 
Before boosting the BTZ solution, it is first appropriate to expand it about
the background of the anti-de~Sitter space. To achieve this, we expand the BTZ
solution (\ref{btz}) to first order in the mass term $8M+1$ and the angular
momentum~$J$. The result is
\begin{equation}
\label{btzfirst1}
ds^2 \approx ds^2_{\rm ads}+(8M+1)dt^2 +\frac{8M+1}{(1+r^2/a^2)^2}dr^2
-8J\,dt\,d\phi.
\end{equation}
Using the coordinates in (\ref{adsco}), the above metric can be rewritten
as
\begin{eqnarray}
\label{btzfirst2}
ds^2&=& ds^2_{\rm ads}+\frac{(8M+1)a^2}{(Z_0^2+Z_3^2)^2}\left
[(Z_3dZ_0-Z_0dZ_3)^2
         +\frac{a^2}{Z_3^2+Z_0^2-a^2)}(Z_3dZ_3+Z_0dZ_0)^2\right] \nonumber\\
    &&-\frac{8Ja}{(Z_3^2+Z_0^2)(Z_3^2+Z_0^2-a^2)}(Z_3dZ_0-Z_0dZ_3)
         (Z_1dZ_2-Z_2dZ_1).
\end{eqnarray}
We now make a Lorentz boost (\ref{lorentz}) in the $Z_1$-direction, rescaling
the mass and angular momentum as
\begin{equation}
\label{rescale}
8M+1=8p\sqrt{1-v^2},\ \ {\rm and}\ \ J=s\sqrt{1-v^2}.
\end{equation}
We then proceed to the ultrarelativistic limit $v\rightarrow1$. In this case,
the two constants $p$ and $s$ can be interpreted physically as the energy and
spin angular momentum of the resulting null particle respectively. It may be
observed that, in this limit, the inequality $J<Ma$ mentioned above is
strictly violated. This is because the mass $M$ and angular momentum $J$ are
rescaled in different ways. However, the limit is still an exact solution even
though it is not strictly the limit of a real rotating black hole. Using this
procedure in (\ref{btzfirst2}), we obtain
\begin{eqnarray}
\label{btzfirst3}
ds^2=&& ds^2_{\rm ads}+ 8\pi p(Z_3-\sqrt{Z_3-a^2})\delta(Z_0+Z_1)(dZ_0+dZ_1)^2
      \nonumber \\
      &&-\frac{8\pi s}{a}\left[Z_2-\frac{Z_2Z_3}{\sqrt{Z_3^2-a^2}}\right]
      \delta(Z_0+Z_1)(dZ_0+dZ_1)^2.
\end{eqnarray}
The two linear terms in the solution (\ref{btzfirst3}) can be removed by the
following discontinuous linear transformation
\begin{eqnarray}
&& Z_2 \rightarrow Z_2 -\frac{4\pi s}{a}U \Theta (U), \nonumber \\
&& Z_3 \rightarrow Z_3 -4\pi p U\Theta (U), \nonumber \\
&& U \rightarrow U, \nonumber \\
&& V \rightarrow V -16\pi ^2 p^2 U\Theta (U) + 8\pi p Z_3 \Theta (U)
+\frac{16\pi^2 s^2}{a^2}U \Theta (U) -\frac{8\pi s}{a}Z_2 \Theta (U),
\end{eqnarray}
where $U=Z_0+Z_1$, $ V=Z_0-Z_1$, and $\Theta$ is the Heaviside step function.
 Therefore, the solution (\ref{btzfirst3}) can be reduced to
\begin{equation}
\label{btzfirst4}
ds^2=ds^2_{\rm ads}-8\pi p|Z_2|\delta(Z_0+Z_1)(dZ_0+dZ_1)^2 
+\frac{8\pi sZ_3}{a}  {\rm sign}(Z_2)\delta(Z_0+Z_1)(dZ_0+dZ_1)^2.
\end{equation}
It is now easy to see that when $s=0$, that is when the angular momentum
vanishes in the original BTZ solution, the solution (\ref{btzfirst4}) is
identical to (\ref{andshock}). Thus, both ultrarelativistic limits of the
DJ solution for $\Lambda<0$ and the spinless BTZ solution are equivalent to
each other. Obviously, the third term in (\ref{btzfirst4}) is the spin effect
of the null particle. In the coordinates (\ref{anticon2}) and (\ref{con}), we
can rewrite  (\ref{btzfirst4}) as
\begin{eqnarray}
\label{btzfirst5}
ds^2&=&ds^2_{\rm ads}+\left( -8\pi p\ a \frac{|\sin\phi|}{\cos^2\phi}
+\frac{8\pi s\ {\rm sign}(\tan \phi)}{\cos\phi|\cos\phi|}
\right)[\delta(\eta-\rho)(d\eta-d\rho)^2 \nonumber \\
 && +\delta(\eta+\rho)(d\eta+d\rho)^2]
\nonumber \\\
&=& ds^2_{\rm ads}+\left( -8\pi p\ a \frac{|\sin\phi|}{\cos^2\phi}
+\frac{8\pi s\ }{\cos^2\phi}{\rm sign}(\sin\phi)
\right)[\delta(\eta-\rho)(d\eta-d\rho)^2 \nonumber \\
 && +\delta(\eta+\rho)(d\eta+d\rho)^2]
\end{eqnarray}
which is clearly identical to (\ref{boloid}) when $s=0$.

\section{Null particles in the DJ background}

In this section we will directly solve the Einstein equations with null
particle sources and re-obtain some of the results of previous sections that
were derived by the boost method. In \cite{DH} Dray and 't~Hooft considered a 
particle moving with the speed of light on the Schwarzschild black hole
horizon, and investigated the back reaction of the particle on the geometry.
In this case, the particle produces an impulsive gravitational wave located on
the Schwarzschild horizon. Loust\'o and S\'anchez \cite{LS} and Sfetsos 
\cite{Sfet} further extended the work of Dray and 't Hooft to nonvacuum
backgrounds and investigated the conditions that should be satisfied when an
impulsive wave is introduced into curved spacetimes. The null particle
solution in  the BTZ background (\ref{btz}) has already been considered in
\cite{Sfet}.

Here we first note that the DJ solution (\ref{dj}) can be rewritten, after
rescaling the coordinate $r$, as
\begin{equation}
\label{dj1}
ds^2=-(1-\Lambda r^2)dt^2 +(1-\Lambda r^2)^{-1}dr^2 +\alpha ^2 r^2d\phi^2.
\end{equation}
Further defining  $\phi'=\alpha \phi$ with $\phi' \in [0, 2\pi\alpha)$,
we have
\begin{equation}
\label{dj2}
ds^2=-(1-\Lambda r^2)dt^2 +(1-\Lambda r^2)^{-1}dr^2 +r^2d\phi'^2,
\end{equation}
which is obviously equivalent to the de~Sitter space locally, but has a
deficit angle $\delta=(1-\alpha)2\pi$. We will first discuss solutions with
null particles located on the cosmological horizon of the de~Sitter case
($\Lambda=1/a^2>0$). The metric (\ref{dj1}) or (\ref{dj2}) also can be
regarded as a hypersurface embedded in the flat spacetime (\ref{desitter})
with
\begin{eqnarray}
&& Z_0=\sqrt{a^2-r^2}\sinh(t/a), \qquad Z_1=r \cos(\alpha \phi), 
    \nonumber\\ 
&& Z_3=\pm \sqrt{a^2-r^2}\cosh(t/a), \qquad Z_2=r\sin(\alpha \phi).
\end{eqnarray}
Introducing the null coordinates
\begin{equation}
u=e^{t/a}F(r), \ \ v=e^{-t/a}F(r),
\end{equation}
where the function $F(r)$ is defined as
\begin{equation}
F(r) \equiv \exp \left(-{1\over a} \int \frac{dr}{(1-r^2/a^2)}\right)
=\left(\frac{a-r}{a+r}\right)^{1/2},
\end{equation}
we can re-express the DJ solution (\ref{dj2}) as
\begin{equation}
\label{dj3}
ds^2=2A(u,v)dudv +r^2(u,v)d\phi'^2,
\end{equation}
where
\begin{equation}
A(u,v)=\frac{(a^2-r^2)}{2\,F^2(r)}, \qquad
r(u,v)=\frac{a(1-uv)}{1+uv}.
\end{equation}
We can now consider the effect of null particles located on the null surface
$u=0$ which is clearly the cosmological horizon $r=a$. In this
three-dimensional theory, this horizon is circular. Following \cite{DH} and
\cite{Sfet}, we can adopt the coordinate shift method on the background
(\ref{dj2}). That is, the following ansatz is employed: For $u<0$ the
spacetime is still the background (\ref{dj3}) while for $u>0$ the spacetime is
(\ref{dj3}) with $v$ shifted as $v\rightarrow v+ f(\phi')$. The function
$f(\phi')$ which will be determined later describes the effect of the sources.
Using this approach, the new solution has the form
\begin{equation}
ds^2=2A(u,v)dudv-2A(u,v)f(\phi')\delta(u)du^2 +r^2(u,v)d\phi'^2,
\end{equation}
which comes from (\ref{dj3}) after making the coordinate shift:
\begin{equation}
u \rightarrow u, \qquad  v\rightarrow v-f(\phi')\Theta (u), \qquad 
\phi'\rightarrow \phi'.
\end{equation}
 For this solution to be consistent with the Einstein equations, the following
conditions must be satisfied at $u=0$ 
\cite{Sfet},
\begin{eqnarray}
\label{con1}
&& A_{,v}=r^2_{,v}=T_{vv}=0, \\
\label{con2}
&& \frac{d^2f}{d\phi'^2}-\frac{r^2_{,uv}}{2A}f
=\frac{8\pi r^2}{A}\tilde{T}_{uu},
\end{eqnarray}
 where $T$ is the energy-momentum tensor of matter generating the DJ geometry,
that is, the cosmological constant and possibly some static point particles,
and $\tilde{T}$ is the energy-momentum tensor of any null particles located on
the surface. Here it should be noticed that the only nonvanishing component of
the energy-momentum tensor for null particles is  $\tilde{T}_{uu}$, and that
this is zero everywhere except at the points where the particles are located.

At the $u=0$ null surface --- that is, on the cosmological horizon 
$r_c=a=1/\sqrt{\Lambda}$ for the DJ ($\Lambda>0$) solution (\ref{dj2}) --- it
is easy to see  that the conditions (\ref{con1}) are satisfied and
\begin{equation}
A(u,v)|_{u=0}=2a^2, \ \ r^2_{,uv}|_{u=0}=-4a^2.
\end{equation}
Then (\ref{con2}) reduces to
\begin{equation}
\label{tbs}
\frac{d^2f}{d\phi'^2}+f=4\pi \tilde{T}_{uu}.
\end{equation}

It may immediately be observed that a solution with $f=4\pi\rho$, where $\rho$
is a constant, represents a uniform distribution of null matter (of density
$\rho$) over the circular horizon. Since equation (\ref{tbs}) is linear, this
component can always be added to other components. However, we will ignore
this possibility in the remainder of this section.

In those parts of the null surface on which $\tilde{T}_{uu}=0$, equation
(\ref{tbs}) has the solution 
\begin{equation}
 f=c\sin(\phi'+\omega')=c\sin\alpha(\phi+\omega)
\end{equation}
 where $c$ and $\omega'=\alpha\omega$ are arbitrary constants. This solution
for $f$ around the circular horizon can always be removed by a discontinuous
coordinate transformation. However, solutions describing several discrete
particles can be constructed by patching different sections of the sine wave,
each with different amplitude and phase. Points at which $f$ is $C^0$ but has
a discontinuous first derivative can be interpreted as points at which null
particles are located. The energy of each particle is then represented by the
jump in the derivative of $f$, and the energy-momentum tensor $\tilde T_{uu}$
is given by a $\delta$-function. On considering (\ref{btzfirst5}), it may be
observed that discontinuities in $f$ may also be permitted. These represent
point particles with spin and, in this case, the energy-momentum tensor
$\tilde T_{uu}$ contains a derivative of a
$\delta$-function.

For example, consider the case in which $n$ particles each of energy $p_i$,
$i=1\dots n$, are located at points $\phi=\phi_i$ around the circular wave.
The solution is then given by 
\begin{equation}
 f=c_i\sin\alpha(\phi+\omega_i) \qquad {\rm for} \qquad
\phi_{i-1}\le\phi\le\phi_i
\end{equation}
 where $n+1\to1$. It is then possible to choose the $2n$ arbitrary constant
$c_i$ and $\omega_i$ such that 
\begin{eqnarray}
&& c_{i+1}\sin\alpha(\phi_i+\omega_{i+1}) -c_i\sin\alpha(\phi_i+\omega_i) =0 
\nonumber\\
&& c_{i+1}\cos\alpha(\phi_i+\omega_{i+1})
-c_i\cos\alpha(\phi_i+\omega_i) ={4\pi p_i\over a}
\end{eqnarray}
 By choosing the constants appropriately, it is possible to construct
solutions in which $n$ ($\ge2$) null particles of arbitrary energy are
distributed arbitrarily round the circular wave. 

In particular, we can consider the two-particle solution in which the
particles are located at opposite ends of a diameter of the circle. Since
$0\le\phi'<2\pi\alpha$ around the circle, we may consider the particles to be
located at points given by $\phi'=0$ and $\phi'=\pi\alpha$. We may also
restrict attention to the case in which the two particles have identical
energy $p$. Such a solution can be constructed by the above method in which
\begin{equation}
c_1=c_2={2\pi p\over a}\,{\rm cosec}\,{\pi\alpha\over2}, \qquad
\omega_1={(1-\alpha)\pi\over2\alpha}, \qquad
\omega_2={(1-3\alpha)\pi\over2\alpha}. 
\end{equation}
This solution represents two null particles propagating in the DJ
($\Lambda>0$) background. In the case in which $\alpha=1$, the background is
the de~Sitter space and the solution can alternatively be written in the form 
\begin{equation}
\label{twonull}
f=\frac{2\pi p}{a}|\sin\phi|.
\end{equation}
This is clearly identical (after some rescaling) to the solution
(\ref{decircle}) that was obtained by boosting two static particles in the
de~Sitter background, and thus confirms this solution.

It may also be observed that, in the particular case in which $\alpha=1/2$,
the deficit angle in $\phi'$ is $\pi$ and a one-particle solution at $\phi=0$
can easily be constructed using 
\begin{equation}
 f=c\sin(\phi/2) \qquad {\rm where} \qquad
0\le\phi\le2\pi. 
\end{equation}

It is also possible to obtain solutions for null particles propagating in the
DJ background with $\Lambda<0$. However, the Dray--'t~Hooft \cite{DH} method
cannot be directly used in this case. Nevertheless, equivalent equations can
be obtained and these will include, for $\alpha=1$, the special cases
(\ref{boloid}) of a null particle and (\ref{btzfirst5}) of a spinning null
particle propagating in an anti-de~Sitter space.

\section{Conclusion and discussion}
We have investigated null particle solutions in the three-dimensional
de~Sitter and anti-de~Sitter spaces by boosting the corresponding static point
source solutions (DJ solutions) \cite{DJ} in the (anti-) de~Sitter
backgrounds. For the de~Sitter case, the resulting solution describes two null
particles located at opposite points on the cosmological horizon which forms a
circle of constant size. For the anti-de~Sitter case, the solution describes a
single null particle located at the point of symmetry of a propagating
hyperbola. We have also boosted the BTZ black hole solution to the
ultrarelativistic limit. Although the BTZ black hole is quite different from
the DJ solution globally, we have found that these two ultrarelativistic
limits are equivalent to each other when the angular momentum of the hole is
zero. This means that the boost method may loose some memory of the original
solution in the process of the boost. In addition, we believe that the angular
momentum of the hole gives the spin effect of the corresponding null particle. 

By using the coordinate shift method, we have also obtained null particle
solutions in DJ background. When $\alpha=1$, the DJ ($\Lambda>0$) solution
reduces to the de~Sitter space, and the results obtained include that derived
using the boost method. It may be observed that the boost method indeed is
very powerful in the derivation of null particle solutions, not only in flat
spacetime, but also in the (anti-) de~Sitter space in three dimensions as well
as four.

Due to the special properties of the geometry in three-dimensional spacetime,
the nature of Einstein gravity is rather different from that in four
dimensions. The static point source solutions, given in \cite{DJH1} in flat
spacetime and in \cite{DJ} in (anti-) de~Sitter space, clearly demonstrate the
differences in the local and global aspects from the four-dimensional
Schwarzschild and Schwarzschild-(anti-)de~Sitter solutions. By comparing the
null particle solution given by Deser and Steif \cite{DS} in three-dimensional
flat spacetime and some results given in this paper, we may observe some
similarities as well as some differences in the null particle solutions in
three and four dimensions. The main difference in four-dimensional spacetime
is that the null particles generate impulsive gravitational waves which are
forbidden in three-dimensional theories. It is of some interest to further
compare spacelike source solutions in three and four-dimensional spacetimes.
Furthermore, it also might be interesting to discuss the geodesics and
particle scattering in the null particle solutions in the (anti-) de~Sitter
space.


\end{document}